\documentclass[11pt,a4paper]{article}
\usepackage[utf8]{inputenc}
\usepackage[T1]{fontenc}
\usepackage{lmodern}
\usepackage[margin=2.6cm]{geometry}
\usepackage{amsmath,amssymb,amsthm}
\usepackage{graphicx}
\usepackage{booktabs}
\usepackage{float}
\usepackage{authblk}
\usepackage{xcolor}
\usepackage{hyperref}
\usepackage{xurl}
\usepackage{natbib}
\floatstyle{ruled}
\newfloat{algorithm}{tbp}{loa}
\floatname{algorithm}{Algorithm}
\hypersetup{colorlinks=true,linkcolor=blue!60!black,citecolor=blue!60!black,urlcolor=blue!60!black}

\graphicspath{{figures/}}

\newtheorem{definition}{Definition}
\newtheorem{proposition}{Proposition}
\newtheorem{theorem}{Theorem}
\newtheorem{corollary}{Corollary}
\newtheorem{lemma}{Lemma}
\theoremstyle{remark}
\newtheorem{remark}{Remark}

\newcommand{\N}{N}
\newcommand{\R}{\mathbb{R}}
\newcommand{\Var}{\operatorname{Var}}
\newcommand{\Cov}{\operatorname{Cov}}

\title{\textbf{Stability Bands for Feature Attribution:\\
The Core, the Nucleolus and the Bargaining Set of Explained-Variance Games}%
\thanks{An earlier version of this paper, titled ``Stability-Certified Explainability
for Economic Forecasting: A Cooperative Game-Theoretic Approach via the Nucleolus
and Bargaining Sets'', was presented at the XXV Jornadas Latinoamericanas de
Teor\'ia Econ\'omica (JOLATE 2026), Universidad de La Punta, San Luis, Argentina,
October 29, 2026. We thank the participants for their comments.}}

\author[1,2]{Fernando Delbianco\thanks{Corresponding author: \href{mailto:fernando.delbianco@uns.edu.ar}{fernando.delbianco@uns.edu.ar}. ORCID: \href{https://orcid.org/0000-0002-1560-2587}{0000-0002-1560-2587}}}
\author[1,2]{Fernando Tohm\'e\thanks{ORCID: \href{https://orcid.org/0000-0003-2988-4519}{0000-0003-2988-4519}}}
\affil[1]{Department of Economics, Universidad Nacional del Sur (UNS), Bah\'ia Blanca, Argentina}
\affil[2]{Institute of Mathematics -- CONICET, Bah\'ia Blanca, Argentina}
\date{}

\begin{document}
\maketitle

\begin{abstract}
Shapley-value attributions are the default way of explaining machine-learning
predictions in economics, but they report a single number per feature and say
nothing about whether that allocation could be contested by a coalition of
features that jointly accounts for more than it receives. We study feature
attribution through the stability concepts of cooperative game theory, namely
the core, the nucleolus and the bargaining set, applied to an
\emph{explained-variance game}, in which the worth of a set of features is the
variance of the model's predictions that those features generate. We report a
\emph{stability band} for every feature (the range of its share over the core,
or over the least core when the core is empty), together with the nucleolus as
the band's distinguished point and a certificate that the reported allocation
admits no justified objection. Three results organise the approach. First, for
linear models the explained-variance game is exactly 2-additive, with pairwise
Harsanyi dividends equal to $2\beta_i\beta_j\sigma_{ij}$, so that
multicollinearity is what creates coalitional structure. Second, in every
2-additive game the prenucleolus coincides with the Shapley value, the
(least) core is point-symmetric around it, and the core is non-empty if and
only if every cut of the interaction graph has non-negative weight. The Shapley
value is therefore always certified to lie in the prebargaining set, but
stability adds information only through the width of the bands and through
interactions of order three or more. Third, for convex games the bands have a
closed form and any proposed allocation can be certified with one minimum cut.
In a hedonic price model the bands show that the usual mean-$|$SHAP$|$
explanation admits a justified objection by the coalition
\{Ratings, ROM, Battery\}. In a gradient-boosting forecaster of retail orders
the exact game has an empty core and strong higher-order interactions; there
the nucleolus departs from the Shapley value by up to 0.19 variance shares, and
pairwise surrogates cannot recover it, whereas sparse third-order surrogates
can.


\end{abstract}

\section{Introduction}

Machine-learning models are now routinely used for forecasting, pricing and
policy evaluation \citep{mullainathan2017machine}, and the institutions that
use them increasingly need to say \emph{why} a model predicts what it does.
The dominant answer is feature attribution through the Shapley value
\citep{shapley1953value}, popularised by SHAP \citep{lundberg2017unified}: a
model is cast as a cooperative game whose players are the features, and each
feature receives its average marginal contribution.

The Shapley value is an allocation rule justified by fairness axioms. It is
not a stability concept. Nothing guarantees that a coalition of features,
say three variables that jointly drive most of the prediction, receives at
least what it could account for on its own. When such a coalition is
under-credited, the explanation can be contested on its own terms: the
coalition has an \emph{objection}. Stability concepts from cooperative game
theory formalise exactly this kind of scrutiny. The \emph{core} contains the
allocations against which no coalition has a positive excess. The
\emph{bargaining set} \citep{aumann1964bargaining,davis1965kernel} contains
those against which every objection can be countered. The \emph{nucleolus}
\citep{schmeidler1969nucleolus} lexicographically minimises the
dissatisfaction of the worst-treated coalitions. Moving from fairness to
stability is in line with the stability principle of veridical data science
\citep{yu2021veridical}, and \citet{yan2021if} have argued for the core as an
attribution concept.

This paper develops that programme for a specific and economically
interpretable game, and is explicit about what stability adds to the Shapley
value and what it does not. Our contributions are the following.

\begin{enumerate}
\item \textbf{An explained-variance game.} The worth of a set of features
$S$ is the variance of the model's predictions when only the features in $S$
vary and the rest are held at a baseline (Section~\ref{sec:game}). Shares of
this game add up to the variance of the predictions. For linear models the game
is exactly 2-additive: its pairwise Harsanyi dividends are
$2\beta_i\beta_j\sigma_{ij}$, so coalitional structure is generated precisely
by multicollinearity (Proposition~\ref{prop:linear}).

\item \textbf{The structure of 2-additive games.} In every game whose
Harsanyi dividends vanish beyond order two, whatever their signs, the
prenucleolus equals the Shapley value, the core and the least core are
point-symmetric around it, the least-core value equals minus one half of the
minimum cut of the interaction graph, and the core is non-empty if and only if
every cut has non-negative weight (Theorem~\ref{thm:2add}). Two consequences
follow. The Shapley value of such a game is always certified to belong to the
prebargaining set. And for pairwise surrogates, a class widely used to make
attribution games tractable, the nucleolus cannot differ from the Shapley
value. Stability then adds information only
through the \emph{bands} around the point and through interactions of order
three or more.

\item \textbf{Stability bands and certificates.} We report, for each
feature, the interval of its share over the core (or over the least core when
the core is empty), with the nucleolus as distinguished point
(Section~\ref{sec:bands}). For convex games the bands have a closed form, and
whether an arbitrary allocation, a SHAP or LIME explanation for instance, lies
in the core (and hence in the bargaining set) is decided by a single minimum
$s$-$t$ cut (Proposition~\ref{prop:mincut}). For general games we give an exact
sequential-LP algorithm for the prenucleolus with a provably correct fixing
rule, and an exact test for justified objections that constructs the objection
vectors explicitly (Section~\ref{sec:algorithms}).

\item \textbf{Two applications.} In a hedonic price model of mobile phones
(Section~\ref{sec:mobile}) the game is exact and convex. The usual
mean-$|$SHAP$|$ explanation turns out to be outside the core: the coalition
\{Ratings, ROM, Battery\} has a justified objection against it. In a gradient
boosting forecaster of retail order quantities (Section~\ref{sec:sales}) we
compute the exact game on all $2^{15}$ coalitions. Its core is empty, its
Harsanyi dividends carry large interactions up to order five, and its
nucleolus departs from the Shapley value by up to $0.19$ variance shares. On
this exact benchmark, pairwise surrogates reach a held-out $R^2$ of about $0.90$
yet miss the nucleolus by about $0.18$, whereas sparse third-order surrogates
recover it to within $0.02$ (Section~\ref{sec:validation}).
\end{enumerate}

All results can be reproduced with the code that accompanies the paper.

\section{Related literature}\label{sec:literature}

\paragraph{Shapley-based attribution.} SHAP \citep{lundberg2017unified} and
TreeSHAP \citep{lundberg2020local} explain individual predictions; global
importance is usually obtained by averaging absolute local attributions.
\citet{sundararajan2020many} show that the choice of game (baseline,
conditional or interventional) matters as much as the choice of solution
concept, and \citet{kumar2020problems} document the difficulties that
correlated features create for Shapley attributions. Global, variance- or
loss-based games have been proposed by \citet{owen2014sobol}, whose Shapley
effects split the variance of a function of independent inputs, and by
\citet{covert2020understanding}, whose SAGE attributes reductions in
predictive loss. In linear regression, variance decompositions of this kind
form the relative-importance literature surveyed by
\citet{gromping2007estimators}. Our explained-variance game belongs to this
family. The difference is that we study its stability concepts rather than
only its Shapley value.

\paragraph{Stability concepts in attribution.} \citet{yan2021if} propose the
core, and specifically the least core, as an alternative to the Shapley value
for data valuation and feature attribution. We add the bargaining set and the
nucleolus, and show that for 2-additive games all of these concepts are
organised around the Shapley value (Theorem~\ref{thm:2add}).

\paragraph{Computation.} Computing the nucleolus requires, in general, a
sequence of linear programs with exponentially many constraints
\citep{kopelowitz1967computation,maschler1979geometric,elkind2009computational};
efficient implementations of this scheme and an independent procedure for
verifying its output are given by \citet{potters1996computing} and
\citet{benedek2021finding}. Polynomial algorithms
exist for structured classes, notably matching games
\citep{kern2003matching,konemann2020computing,benedek2023complexity}, and,
more generally, whenever the minimum-excess (separation) problem can be solved
in polynomial time \citep{faigle2001computation}. For convex games this
yields a polynomial algorithm based on the ellipsoid method. A combinatorial,
strongly polynomial algorithm has recently been proposed by
\citet{maggiorano2025strongly}, but its correctness has been disputed by
\citet{meinhardt2025polynomial}, so we rely only on \citet{faigle2001computation}. Games given by Harsanyi
dividends on a graph (induced-subgraph games) were studied by
\citet{deng1994complexity}: the Shapley value is easy to compute, while core
questions are hard when weights may be negative. We use their framework, the
$k$-additive representation of \citet{grabisch1997k}, the theory of convex
games \citep{shapley1971cores,maschler1972kernel,ichiishi1981supermodularity}
and the selection (maximum-closure) problem of \citet{rhys1970selection},
\citet{balinski1970selection} and \citet{picard1976maximal}.

\paragraph{Evaluation of explanations.} Following the assessment framework of
\citet{calzarossa2025assessment}, we evaluate explanations by robustness
(bootstrap stability) and complexity (concentration), in addition to
deletion-based faithfulness \citep{samek2017evaluating}. Following
\citet{chen2024measuring}, we also illustrate how stability-based attributions
can feed a group-comparison diagnostic.

\section{Cooperative games: definitions and known results}\label{sec:cgt}

A TU game is a pair $(\N,v)$ with $\N=\{1,\dots,n\}$ and $v:2^\N\to\R$,
$v(\emptyset)=0$. For $x\in\R^n$ write $x(S)=\sum_{i\in S}x_i$. A
\emph{preimputation} is an $x$ with $x(\N)=v(\N)$. In attribution problems
shares may legitimately be negative, so we work with preimputations
throughout. The \emph{excess} of $S$ at $x$ is $e(S,x)=v(S)-x(S)$, the amount
by which $S$ is under-credited.

\begin{definition}[Core, least core, prenucleolus]
The \emph{$\varepsilon$-core} is
\[
C_\varepsilon(v)=\{x\in\R^n:\ x(\N)=v(\N),\ e(S,x)\le\varepsilon\ \text{for all}\ \emptyset\neq S\subsetneq \N\};
\]
the \emph{core} is $C_0(v)$. The \emph{least-core value} is
$\varepsilon^*(v)=\min\{\varepsilon: C_\varepsilon(v)\neq\emptyset\}$, and
$C_{\varepsilon^*}(v)$ is the \emph{least core}; the core is non-empty iff
$\varepsilon^*\le 0$. The \emph{prenucleolus} is the unique preimputation whose
vector of excesses, arranged in non-increasing order, is lexicographically
minimal \citep{schmeidler1969nucleolus,peleg2007introduction}. It lies in the
least core, hence in the core whenever the latter is non-empty.
\end{definition}

\begin{definition}[Objections and the (pre)bargaining set]
Let $x$ be a preimputation and $i\neq j$. An \emph{objection} of $i$ against
$j$ is a pair $(S,y)$ with $i\in S$, $j\notin S$, $y\in\R^S$, $y(S)=v(S)$ and
$y_k>x_k$ for all $k\in S$. A \emph{counter-objection} to it is a pair $(T,z)$
with $j\in T$, $i\notin T$, $z\in\R^T$, $z(T)=v(T)$, $z_k\ge y_k$ for
$k\in T\cap S$ and $z_k\ge x_k$ for $k\in T\setminus S$. An objection is
\emph{justified} if it has no counter-objection. The \emph{prebargaining set}
$\mathcal{PM}(v)$ is the set of preimputations against which no justified
objection exists \citep{davis1965kernel,peleg2007introduction}. Restricting
$x$ to individually rational preimputations gives the classical bargaining
set $\mathcal{M}^{(i)}_1$ of \citet{aumann1964bargaining}.
\end{definition}

\begin{definition}[Harsanyi dividends, $k$-additivity, convexity]
Every game has a unique representation $v(S)=\sum_{T\subseteq S} d(T)$, where
$d(T)=\sum_{R\subseteq T}(-1)^{|T|-|R|}v(R)$ are the Harsanyi dividends
\citep{harsanyi1963simplified}. The game is \emph{$k$-additive} if $d(T)=0$
whenever $|T|>k$ \citep{grabisch1997k}. It is \emph{convex} if
$v(S\cup T)+v(S\cap T)\ge v(S)+v(T)$ for all $S,T$.
\end{definition}

We will use the following known facts.

\begin{proposition}[Known results]\label{prop:known}
\begin{enumerate}
\item[(a)] The Shapley value is $\phi_i(v)=\sum_{T\ni i} d(T)/|T|$.
\item[(b)] \emph{(Kohlberg's criterion.)} A preimputation $x$ is the
prenucleolus iff for every $\alpha\in\R$ the collection
$\mathcal D(\alpha,x)=\{\emptyset\ne S\subsetneq \N: e(S,x)\ge\alpha\}$ is
empty or balanced \citep{kohlberg1971nucleolus,peleg2007introduction}.
\item[(c)] $C_0(v)\subseteq\mathcal{PM}(v)$, and the prenucleolus belongs to
the prekernel, which is contained in $\mathcal{PM}(v)$
\citep{davis1965kernel,peleg2007introduction}. Hence the prenucleolus always
belongs to the prebargaining set.
\item[(d)] If all dividends of order $\ge2$ are non-negative, $v$ is convex.
If $v$ is convex, the core is non-empty and equals the convex hull of the
marginal vectors, and the Shapley value lies in it
\citep{shapley1971cores,ichiishi1981supermodularity}. Moreover, the bargaining
set coincides with the core and the kernel with the nucleolus
\citep{maschler1972kernel}.
\end{enumerate}
\end{proposition}

The first claim of (d) has a direct proof. Singleton dividends are additive and
do not affect convexity, so we may assume $d(T)\ge0$ for every $T$. Then
\[
v(S\cup T)+v(S\cap T)-v(S)-v(T)=\sum_{R}d(R)\big(\mathbb 1[R\subseteq S\cup T]+\mathbb 1[R\subseteq S\cap T]-\mathbb 1[R\subseteq S]-\mathbb 1[R\subseteq T]\big).
\]
For each $R$ the bracket equals $1$ when $R\subseteq S\cup T$ but
$R\not\subseteq S$ and $R\not\subseteq T$, and $0$ in every other case. Hence
the sum is non-negative.

Part (d) has a simple consequence that we use repeatedly. Every marginal vector
assigns player $i$ an amount between $v(\{i\})$ (when $i$ comes first) and
$v(\N)-v(\N\setminus\{i\})$ (when $i$ comes last), and supermodularity makes
these the extremes. Hence, in a convex game,
\begin{equation}\label{eq:convexband}
\min_{x\in C_0(v)} x_i=v(\{i\}),\qquad \max_{x\in C_0(v)}x_i=v(\N)-v(\N\setminus\{i\}).
\end{equation}

\section{The explained-variance game}\label{sec:game}

Let $f:\R^n\to\R$ be a fitted model, $\mathcal X=\{x^{(1)},\dots,x^{(m)}\}$ an
evaluation sample and $b\in\R^n$ a baseline, for instance the training mean.
For $S\subseteq\N$ let $f_S(x)=f(x_S,b_{\N\setminus S})$ denote the prediction
when only the features in $S$ take their observed values.

\begin{definition}[Explained-variance game]
$v(S)=\Var_{x\sim\mathcal X}\big[f_S(x)\big]$.
\end{definition}

Then $v(\emptyset)=0$, and $v(\N)$ is the variance of the model's predictions
over $\mathcal X$. We always report the normalised game $v/v(\N)$, so that
attributions are \emph{shares of the prediction variance} that add up to one.
A feature that the model ignores is a null player and receives zero in every
solution considered here. Negative shares are possible. They identify features
whose variation offsets that of others, which in the economic examples below
is typically a sign of substitution. The game is \emph{global}: it explains
the model over a population rather than a single prediction, as do the
analyses of \citet{owen2014sobol} and \citet{covert2020understanding}. A local
version is obtained with $\mathcal X$ a neighbourhood of a point.

\begin{proposition}[Linear models]\label{prop:linear}
Let $f(x)=\beta_0+\beta^\top x$ and let $\Sigma=(\sigma_{ij})$ be the
covariance matrix of $\mathcal X$. Then $v(S)=\sum_{i,j\in S}\beta_i\beta_j\sigma_{ij}$,
and $v$ is 2-additive with
\[
d(\{i\})=\beta_i^2\sigma_{ii},\qquad d(\{i,j\})=2\beta_i\beta_j\sigma_{ij}.
\]
Its Shapley value is $\phi_i=\beta_i\Cov(x_i,f(x))$, and $v$ is convex if and
only if $\beta_i\beta_j\sigma_{ij}\ge0$ for all $i\neq j$.
\end{proposition}

\begin{proof}
$f_S(x)-\mathbb E f_S=\sum_{i\in S}\beta_i(x_i-\bar x_i)$, whose variance is
the stated quadratic form. A quadratic form restricted to $S$ is the sum of its
diagonal terms over $i\in S$ and its off-diagonal pairs over $\{i,j\}\subseteq S$.
By uniqueness of the Harsanyi representation these are the dividends, and all
dividends of order $\ge3$ vanish. Proposition~\ref{prop:known}(a) gives
$\phi_i=\beta_i^2\sigma_{ii}+\sum_{j\ne i}\beta_i\beta_j\sigma_{ij}=\beta_i\Cov(x_i,\beta^\top x)$.
For a 2-additive game,
$v(S\cup\{i\})-v(S)=d(\{i\})+\sum_{j\in S}d(\{i,j\})$, which is non-decreasing
in $S$ for every $i$ iff all $d(\{i,j\})\ge0$.
\end{proof}

The interaction structure of a linear model's explanation is therefore
entirely produced by correlation among regressors. Two features whose
contributions $\beta_i x_i$ and $\beta_j x_j$ co-move \emph{positively} are
\emph{complements} in the game ($d(\{i,j\})>0$), and two whose contributions
co-move negatively are \emph{substitutes} ($d(\{i,j\})<0$). Substitution is
what can empty the core, as we show next. The share
$\phi_i=\beta_i\Cov(x_i,\hat y)/\Var(\hat y)$ is a familiar variance
decomposition \citep{gromping2007estimators}. We now show what stability adds
to it.

\section{Stability bands and the structure of 2-additive games}\label{sec:bands}

\begin{definition}[Stability band]
The \emph{stability band} of feature $i$ is
$B_i=[\min_{x\in C}x_i,\ \max_{x\in C}x_i]$, where $C=C_0(v)$ if the core is
non-empty and $C=C_{\varepsilon^*}(v)$ (the least core) otherwise. We report
$B_i$ together with the nucleolus, which always lies in $C$.
\end{definition}

A wide band means that the data support many stable allocations of credit to
that feature. A narrow band means that stability pins the share down. When the
core is empty, the bands describe the allocations that minimise the largest
complaint; $\varepsilon^*$ measures how large that complaint must be.

For a 2-additive game write $w_i=d(\{i\})$, $w_{ij}=d(\{i,j\})$ (of any sign),
and for $\emptyset\ne S\subsetneq\N$ let
$c(S)=\sum_{i\in S,\,j\notin S}w_{ij}$ be the weight of the cut $(S,\N\setminus S)$
in the interaction graph.

\begin{theorem}[2-additive games]\label{thm:2add}
Let $v$ be 2-additive with Shapley value $\phi$, so that
$\phi_i=w_i+\tfrac12\sum_{j\ne i}w_{ij}$. Then:
\begin{enumerate}
\item[(i)] $e(S,\phi)=-\tfrac12 c(S)=e(\N\setminus S,\phi)$ for every
$\emptyset\neq S\subsetneq \N$.
\item[(ii)] $\phi$ is the prenucleolus of $v$.
\item[(iii)] $\varepsilon^*(v)=-\tfrac12\min_{S}c(S)$. Hence the core is
non-empty iff every cut has non-negative weight, iff $\phi\in C_0(v)$.
\item[(iv)] For every $\varepsilon$, $C_\varepsilon(v)$ is point-symmetric
around $\phi$: $x\in C_\varepsilon(v)\Rightarrow 2\phi-x\in C_\varepsilon(v)$.
In particular, every stability band is centred at $\phi_i$.
\item[(v)] $\phi\in\mathcal{PM}(v)$. If all $w_{ij}\ge0$, then $B_i=[w_i,\ w_i+\sum_{j\ne i}w_{ij}]$
and the bands are bands of the bargaining set.
\end{enumerate}
\end{theorem}

\begin{proof}
Write $W(S)=\sum_{\{i,j\}\subseteq S}w_{ij}$. Then $v(S)=\sum_{i\in S}w_i+W(S)$, and
$v(\N)=v(S)+v(\N\setminus S)+c(S)$ for every $S$.

(i) $\phi(S)=\sum_{i\in S}w_i+\tfrac12\sum_{i\in S}\sum_{j\neq i}w_{ij}
=\sum_{i\in S}w_i+W(S)+\tfrac12c(S)=v(S)+\tfrac12c(S)$, since each pair inside
$S$ is counted twice and each pair across the cut once. Hence
$e(S,\phi)=-\tfrac12c(S)$, and $c(S)=c(\N\setminus S)$.

(ii) By (i), for every $\alpha$ the collection $\mathcal D(\alpha,\phi)$ is
closed under complementation. If it is non-empty it consists of $q\ge1$ pairs
$\{S,\N\setminus S\}$. Giving weight $1/q$ to each member, every player is
covered exactly once by each pair, so the weights sum to one for every player
and the collection is balanced. By Proposition~\ref{prop:known}(b), $\phi$ is the
prenucleolus.

(iii) For any preimputation $x$,
$e(S,x)+e(\N\setminus S,x)=v(S)+v(\N\setminus S)-v(\N)=-c(S)$, so
$\max\{e(S,x),e(\N\setminus S,x)\}\ge-\tfrac12c(S)$ and
$\max_S e(S,x)\ge-\tfrac12\min_S c(S)$. By (i), $\phi$ attains this bound, so it is
the value of $\varepsilon^*$. The core is non-empty iff $\varepsilon^*\le0$,
iff $\min_S c(S)\ge0$, and in that case $\phi$ attains $\varepsilon^*\le0$.

(iv) Let $x\in C_\varepsilon$ and $x'=2\phi-x$. Using (i),
$e(S,x')=v(S)-2\phi(S)+x(S)=-v(S)-c(S)+x(S)$, while
$e(\N\setminus S,x)=v(\N\setminus S)-v(\N)+x(S)=-v(S)-c(S)+x(S)$. Hence
$e(S,x')=e(\N\setminus S,x)\le\varepsilon$ for all $S$, so $x'\in C_\varepsilon$.
A set that is symmetric around $\phi$ has coordinate ranges centred at $\phi$.

(v) The first claim follows from (ii) and Proposition~\ref{prop:known}(c). If
$w_{ij}\ge0$, the game is convex (Proposition~\ref{prop:linear}), and
\eqref{eq:convexband} gives $\min x_i=v(\{i\})=w_i$ and
$\max x_i=v(\N)-v(\N\setminus\{i\})=w_i+\sum_{j\ne i}w_{ij}$. By
Proposition~\ref{prop:known}(d) the core coincides with the bargaining set.
\end{proof}

\begin{corollary}\label{cor:linear}
For a linear model, the variance share $\beta_i\Cov(x_i,\hat y)/\Var(\hat y)$
is the prenucleolus of the explained-variance game (and its nucleolus when
the game is convex) and the centre of its stability bands, and it belongs to the prebargaining set. If all
$\beta_i\beta_j\sigma_{ij}\ge0$, the band of feature $i$ runs from its
stand-alone share $\beta_i^2\sigma_{ii}/\Var(\hat y)$ to its stand-alone share
plus all of its interaction shares.
\end{corollary}

Theorem~\ref{thm:2add} has a sobering implication for attribution methods
built on pairwise surrogates $g(S)=w_0+\sum_{i\in S}w_i+\sum_{\{i,j\}\subseteq S}w_{ij}$:
whatever the signs of the weights, the nucleolus of such a surrogate
\emph{is} its Shapley value. Stability concepts can add information to a
pairwise explanation in only two ways: through the bands, which quantify how
much credit could be moved between features without creating a justified
complaint, and through certification of \emph{other} allocations, such as
mean-$|$SHAP$|$ or LIME shares, which need not be stable. To obtain a point
attribution that genuinely differs from the Shapley value, the game must have
interactions of order three or higher. We show in Section~\ref{sec:sales} that
this is the case for a gradient-boosting forecaster, and that it matters.

A related caveat concerns computation. Pairwise surrogates are sometimes
presented as (fractional) matching games in order to borrow the polynomial
nucleolus algorithms available for that class
\citep{kern2003matching,konemann2020computing}. They are not matching games.
The worth of a coalition in a matching game is the value of an optimal
(fractional) matching inside it, which involves vertex-capacity constraints,
whereas a pairwise surrogate simply adds up every vertex and edge weight. A
unit-weight triangle is worth $3/2$ as a fractional matching game and $3$ as a
pairwise surrogate. The algorithms we use instead are those of
Section~\ref{sec:algorithms}.

Part (iii) also shows where hardness enters. For signed weights, deciding
whether the core is empty amounts to a minimum-cut problem with negative edges,
which contains maximum cut. This is consistent with the NP-completeness results
of \citet{deng1994complexity}. With non-negative interactions, by contrast, the
core is never empty.

\section{Algorithms}\label{sec:algorithms}

\subsection{The prenucleolus}

Algorithm~\ref{alg:nucleolus} is the sequential linear programme of
\citet{maschler1979geometric} with two implementation details that are
essential for correctness: which coalitions are fixed at each stage, and when
to stop.

\begin{algorithm}[H]
\caption{Prenucleolus by sequential linear programming}\label{alg:nucleolus}
\textbf{Input:} $n$ and the values $v(S)$ of all coalitions (or, for convex
games, a separation oracle; Section~\ref{sec:convexalg}).\\
\textbf{Output:} the prenucleolus $x^*$.
\begin{enumerate}
\setlength{\itemsep}{1pt}
\item Set $\mathcal E\gets\{x(\N)=v(\N)\}$ (equality system); all proper coalitions are \emph{free}.
\item For $k=1,2,\dots$:
  \begin{enumerate}
  \item Solve $\min\varepsilon$ subject to $\mathcal E$ and $v(S)-x(S)\le\varepsilon$
        for every free $S$. Let $\varepsilon_k$ be the optimal value and
        $\lambda_S\ge0$ the optimal dual multipliers.
  \item Fix every free $S$ with $\lambda_S>0$: add $v(S)-x(S)=\varepsilon_k$ to $\mathcal E$.
  \item Discard every free $S$ whose indicator vector lies in the span of the
        rows of $\mathcal E$.
  \item If $\operatorname{rank}(\mathcal E)=n$, return the unique solution $x^*$ of $\mathcal E$.
  \end{enumerate}
\end{enumerate}
\end{algorithm}

\begin{lemma}\label{lem:alg1}
Algorithm~\ref{alg:nucleolus} returns the prenucleolus after at most $n-1$
stages.
\end{lemma}

\begin{proof}
The duals of the constraints $v(S)-x(S)\le\varepsilon$ sum to one (this is the
dual constraint of $\varepsilon$), so some $\lambda_S>0$ and at least one coalition
is fixed per stage. By complementary slackness, a coalition with a positive
dual in \emph{some} optimal dual solution is tight at \emph{every} optimal primal
solution. Adding its equality therefore leaves the optimal face of stage $k$
unchanged. A coalition whose indicator lies in the span of $\mathcal E$ has
constant excess on the current face and can no longer affect the
lexicographic order. Every fixed coalition is therefore linearly independent
of $\mathcal E$, the rank grows by at least one per stage, and the procedure
stops after at most $n-1$ stages. The remaining argument is the standard
proof of the sequential-LP characterisation of the (pre)nucleolus
\citep{maschler1979geometric}.
\end{proof}

\begin{remark}
The fixing rule matters. Fixing every coalition that happens to be tight at
the particular vertex returned by the solver can fix coalitions that are
\emph{not} tight on the whole optimal face, and the output is then not the
nucleolus. Take the five-player game with unit dividends on the pairs
$\{1,2\},\{1,3\},\{1,4\},\{1,5\},\{2,3\},\{2,4\}$, and the four-player game with
unit dividends on $\{1,2\}$ and $\{2,3\}$. On these games such a rule can return
$(2,0,1,2,1)$ and $(1,0,1,0)$, whereas Algorithm~\ref{alg:nucleolus} returns
the correct $(2,1.5,1,1,0.5)$ and $(0.5,1,0.5,0)$. Both games are 2-additive,
and the correct outputs coincide with the Shapley value, as
Theorem~\ref{thm:2add}(ii) requires.

The rule in step 2(b) is one of several correct ones. A coalition can
equivalently be fixed when its payoff $x(S)$ is constant on the whole optimal
face, which is checked with two auxiliary LPs (its minimum and maximum over
the face) \citep{kopelowitz1967computation}, or with the linear-algebraic
test of \citet{potters1996computing}. The replication code includes an
independent implementation of the two-LP rule as a regression test. On the
two games above, on 30 random games with three to five players and signed
dividends up to order three, and on 30 random 2-additive games, the two
implementations agree. The output can also be checked against the
balancedness criterion of \citet{benedek2021finding}.
\end{remark}

\begin{remark}
Sparsity of the dividend representation does not reduce the number of
coalition constraints in the LPs of Algorithm~\ref{alg:nucleolus}; it only
lowers the cost of evaluating their right-hand sides, to $O(n|\mathcal K|)$ per
coalition. What non-negative sparse dividends buy is convexity, and with it the
oracle of Section~\ref{sec:convexalg}.
\end{remark}

\subsection{Convex games: closed-form bands and minimum-cut certificates}\label{sec:convexalg}

Suppose the game is given by a sparse family $\mathcal K$ of non-zero
dividends, all non-negative for $|T|\ge2$ (a convex game). Then the bands
follow from \eqref{eq:convexband} at cost $O(n|\mathcal K|)$. It remains to decide
whether a \emph{proposed} allocation $x$ (a SHAP, LIME or analyst-supplied
explanation) is stable, that is, whether $\max_S e(S,x)\le0$.

\begin{proposition}[Minimum-cut certificate]\label{prop:mincut}
Let $v(S)=\sum_{T\in\mathcal K,\,T\subseteq S}d(T)$ with $d(T)\ge0$ for $|T|\ge2$.
Then $\max_{S}\big[v(S)-x(S)\big]$ equals the optimal value of a
maximum-closure problem and can be computed with one minimum $s$-$t$ cut in a
network with $n+|\mathcal K|+2$ nodes. The cut also returns a coalition
attaining the maximum.
\end{proposition}

\begin{proof}
Selecting a hyperedge $T$ (with $|T|\ge2$) earns $d(T)\ge0$ and requires
selecting all its members. Selecting player $i$ earns $g_i=d(\{i\})-x_i$, of
either sign. The total earning of a closed selection $S$ is exactly
$v(S)-x(S)$. This is the selection problem of \citet{rhys1970selection} and
\citet{balinski1970selection}, which reduces to a minimum cut
\citep{picard1976maximal}: connect the source to each item with positive
earning (capacity equal to the earning), each item with negative earning to
the sink, and each hyperedge to its members with infinite capacity. The
maximum equals the sum of positive earnings minus the minimum cut, and the
source side of the cut is an optimal $S$.
\end{proof}

Since the core equals the bargaining set in a convex game, a non-positive cut
value certifies that $x$ admits no justified objection. A positive value
exhibits the objecting coalition. The same oracle separates the core
constraints, so core LPs can be solved by cutting planes without enumerating
coalitions; and since the minimum-excess problem is polynomial, the nucleolus of
such games is polynomial \citep{faigle2001computation}.

\subsection{General games: exact enumeration and an objection test}

For general games (non-convex, or not $k$-additive) we compute $v$ on all
$2^n$ coalitions when $n$ is moderate, as in our 15-feature application, and
use Algorithm~\ref{alg:nucleolus}, $2n$ LPs for the bands, and one LP for
$\varepsilon^*$. To test whether a given $x$ lies in $\mathcal{PM}(v)$ we search
for a justified objection directly. For a pair $(i,j)$ and a coalition $S$
with $i\in S$, $j\notin S$, a counter-objection through $T$ exists iff
$v(T)\ge y(T\cap S)+x(T\setminus S)$. Hence an objection through $S$ is
justified iff the LP
\begin{equation}\label{eq:objlp}
\max\ \delta\quad\text{s.t.}\quad y(S)=v(S),\ \ y_k\ge x_k+\delta\ (k\in S),\ \
y(T\cap S)+x(T\setminus S)\ge v(T)+\delta\ \ (\forall T\ni j,\ T\not\ni i)
\end{equation}
has $\delta^*>0$, in which case its solution $y$ is an explicit justified
objection. The test needs $O(n^2 2^n)$ such LPs, which is practical for
$n\lesssim 10$. For larger $n$ we rely on
Proposition~\ref{prop:known}(c): the prenucleolus is always in
$\mathcal{PM}(v)$, so no test is needed for it.

\subsection{Sparse $k$-additive surrogates}\label{sec:surrogates}

When $2^n$ evaluations of $v$ are too expensive, we fit a $k$-additive
surrogate $\hat v(S)=\sum_{T\subseteq S,\,|T|\le k}\hat d(T)$ by LASSO
regression of $v(S)$ on the indicators $\mathbb 1[T\subseteq S]$ over a random
sample of coalitions. Its fidelity should be judged at two levels: the
held-out $R^2$ of $\hat v$, and, more importantly, the distance between the
solution computed on $\hat v$ and the solution of $v$. Theorem~\ref{thm:2add}
predicts that a 2-additive surrogate can match the Shapley value of $v$ but
not its nucleolus whenever $v$ has substantial higher-order dividends.

One might also constrain the surrogate's dividends of order $\ge2$ to be
non-negative, so that $\hat v$ is convex and the certificates of
Section~\ref{sec:convexalg} apply. Those certificates then concern $\hat v$,
not $v$, and Section~\ref{sec:validation} shows that the cost can be severe. Making
the coefficients non-negative by adding a constant to all of them is not an
alternative. Unlike adding an additive vector, it is not a strategic
equivalence: it adds that constant times the number of active simplices
contained in $S$ to each $v(S)$, which changes the game, its core and its
nucleolus.

\begin{table}[H]
\centering
\caption{Computational profile ($n$ features, $|\mathcal K|$ non-zero dividends).}
\label{tab:complexity}
\small
\begin{tabular}{p{3.6cm}p{5.0cm}p{5.4cm}}
\toprule
Task & Convex game, sparse dividends & General game \\
\midrule
Shapley value & $O(\sum_{T\in\mathcal K}|T|)$ & $O(n2^n)$ (M\"obius transform) \\
Stability bands & closed form \eqref{eq:convexband}, $O(n|\mathcal K|)$ & $2n$ LPs with $2^n$ rows \\
Certify $x$ (core; bargaining set) & one min cut on $n+|\mathcal K|+2$ nodes & core: $O(2^n)$; $\mathcal{PM}$: $O(n^22^n)$ LPs \\
Prenucleolus & polynomial (separation oracle) & at most $n-1$ LPs with $2^n$ rows \\
\bottomrule
\end{tabular}
\end{table}

\section{Illustrations}\label{sec:illustrations}

\subsection{A hedonic price model for mobile phones}\label{sec:mobile}

The data contain $807$ mobile phones with their price in Indian rupees and
seven characteristics: customer rating, RAM, internal storage (ROM), screen
size, primary and front camera resolution, and battery capacity (Kaggle;
\url{https://bit.ly/4kCyJ3O}). We estimate a hedonic regression of log price
on the standardised characteristics by LASSO with five-fold cross-validated
penalty ($\alpha=0.028$; in-sample $R^2=0.671$, cross-validated $R^2=0.662$).
RAM and screen size are excluded by the penalty and are therefore null players.

By Proposition~\ref{prop:linear} the explained-variance game is exactly
2-additive, and no surrogate is needed. The dividends computed from the
coefficients and the sample covariance agree with a direct evaluation of
$v(S)$ on all coalitions up to $7\times10^{-16}$. All pairwise dividends are
non-negative. Both camera coefficients are negative, and both cameras are
negatively correlated with ratings, storage and battery, whose coefficients
are positive, so every pair of contributions $\beta_ix_i$, $\beta_jx_j$
co-moves positively. The game is therefore convex. The largest
interactions are Ratings--ROM ($0.186$ of the prediction variance),
Ratings--Battery ($0.120$) and ROM--Battery ($0.080$).

\begin{table}[H]
\centering
\caption{Mobile prices: shares of the variance of predicted log price. The
nucleolus equals the Shapley value of $v$ (Corollary~\ref{cor:linear}); the
core band of feature $i$ is $[w_i, w_i+\sum_j w_{ij}]$ and coincides with the
bargaining-set band. SHAP and LIME columns are mean absolute attributions
rescaled to sum to one.}
\label{tab:mobile}
\begin{tabular}{lrrrcrr}
\toprule
Feature & $\beta$ & $w_i$ & Nucleolus & Core = bargaining-set band & SHAP & LIME \\
\midrule
Ratings       & 0.706 & 0.300 & 0.466 & [0.300, 0.633] & 0.425 & 0.382 \\
ROM           & 0.516 & 0.160 & 0.300 & [0.160, 0.439] & 0.259 & 0.280 \\
Battery power & 0.314 & 0.059 & 0.163 & [0.059, 0.267] & 0.172 & 0.170 \\
Primary camera & $-0.274$ & 0.045 & 0.067 & [0.045, 0.089] & 0.126 & 0.148 \\
Front camera  & $-0.035$ & 0.001 & 0.004 & [0.001, 0.007] & 0.019 & 0.019 \\
Screen size   & 0 & 0 & 0 & [0, 0] & 0 & 0 \\
RAM           & 0 & 0 & 0 & [0, 0] & 0 & 0 \\
\bottomrule
\end{tabular}
\end{table}

\begin{figure}[H]
\centering
\includegraphics[width=0.85\linewidth]{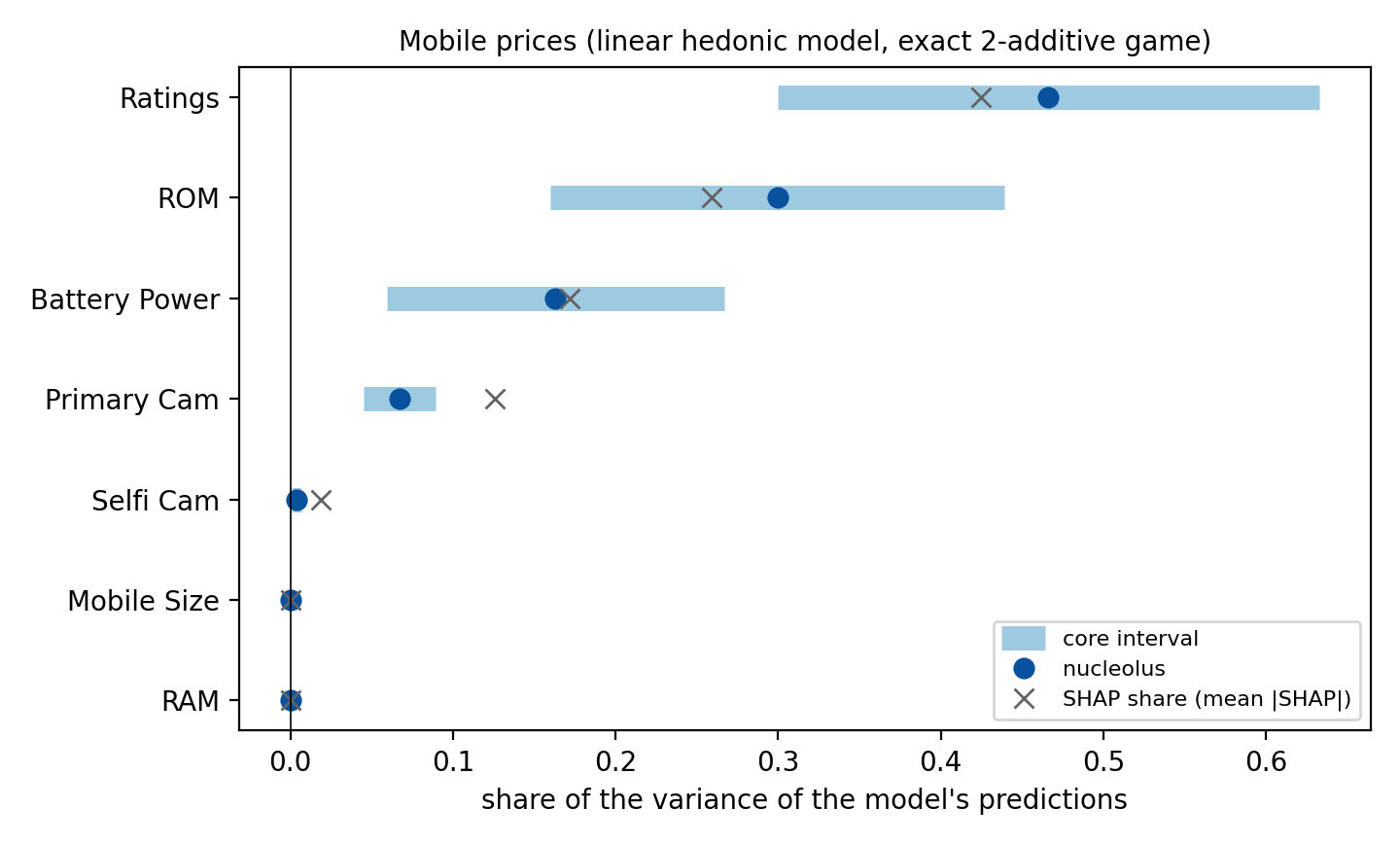}
\caption{Mobile prices: core (= bargaining-set) bands, nucleolus (equal here
to the Shapley value of $v$) and mean-$|$SHAP$|$ shares. The SHAP share of the
primary camera lies outside its band.}
\label{fig:mobile}
\end{figure}

Table~\ref{tab:mobile} and Figure~\ref{fig:mobile} report the results. The
bands computed from \eqref{eq:convexband}, by $2n$ LPs on all coalitions, and
by cutting planes with the minimum-cut oracle (27 cuts) coincide up to
$10^{-12}$. The nucleolus sits exactly at the centre of every band, as
Theorem~\ref{thm:2add}(iv) requires, and an exact search with \eqref{eq:objlp}
finds no justified objection against it.

Two features of the bands are informative. First, they are wide for the three
complementary characteristics: Ratings can be credited with anything between
$30\%$ and $63\%$ of the prediction variance without giving any coalition a
complaint, because the share it holds jointly with ROM and Battery can be
allocated in many stable ways. Second, the band of the primary camera,
$[0.045,0.089]$, is narrow, and the conventional explanation lies outside it.
Mean-$|$SHAP$|$ assigns the camera $0.126$ and mean-$|$LIME$|$ assigns it $0.148$;
in exchange, both under-credit the complementary block \{Ratings, ROM,
Battery\}. The minimum-cut certificate shows that the SHAP allocation leaves a
coalition with excess $0.050$ of the prediction variance, and the LP
\eqref{eq:objlp} returns an explicit justified objection of Ratings against
the primary camera through \{Ratings, ROM, Battery\}. The LIME allocation has
maximum excess $0.073$ and a justified objection through the same coalition.
In this precise sense, the usual global SHAP and LIME explanations of this
model are not stable, while the nucleolus and every point in the bands are.

\subsection{A gradient-boosting forecaster of retail orders}\label{sec:sales}

The second application uses the US Regional Sales data (Kaggle; $7{,}991$
orders, 2018--2020). The target is the order quantity and there are 15
features: sales team, customer, store and product identifiers; discount,
unit cost, unit price and margin; sales channel and warehouse; month and day
of week of the order; and three lead times (order to shipment, order to
delivery, procurement to order). A gradient-boosting regressor (300 trees of depth 4) is trained on the first
$70\%$ of orders in time and evaluated on the remaining $30\%$. We stress at
the outset that the forecaster has no out-of-sample skill (training $R^2=0.274$,
test $R^2=-0.033$): order quantities in these data are essentially
unpredictable from the available features. The attributions below therefore
describe how the \emph{fitted model} distributes the variance of its
forecasts, not the data-generating process. The model is a useful test case
because it is a genuinely non-additive function of 15 inputs, but its
explanations should not be read as economic findings about order behaviour.

Because $2^{15}=32{,}768$ coalitions are few enough to evaluate, we compute
the \emph{exact} game on $500$ test-period orders, with the training mean as
baseline. On a standard laptop-class machine this takes under a minute; the
prenucleolus then takes about three seconds and the bands about six. Figure~\ref{fig:fidelity}(a) summarises the Harsanyi
dividends by order. The game is far from 2-additive: first-order dividends add
up to $2.39$ and second-order ones to $-2.31$ times $v(\N)$, and the absolute
mass of order 3, 4 and 5 interactions ($3.70$, $3.00$ and $1.99$) is as large
as that of the pairs ($3.55$). A quarter ($25.7\%$) of the dividends of order
two or more are negative, the game is not convex, and its core is \emph{empty}:
$\varepsilon^*=0.608$, so every allocation leaves some coalition at least $61\%$
of $v(\N)$ short. A natural source of substitution is that unit price, unit
cost and margin are linearly dependent, so the model can route the same
information through any two of them.

\begin{table}[H]
\centering
\caption{Retail forecaster: shares of the variance of the forecasts (exact
game). Bands are least-core intervals ($\varepsilon^*=0.608$). TreeSHAP
column: mean absolute TreeSHAP values rescaled to sum to one.}
\label{tab:sales}
\begin{tabular}{lrcrr}
\toprule
Feature & Nucleolus & Least-core band & Shapley value of $v$ & TreeSHAP \\
\midrule
Unit cost          & $+0.434$ & [$+0.395$, $+0.496$] & $+0.322$ & 0.113 \\
Store ID           & $+0.201$ & [$+0.153$, $+0.218$] & $+0.155$ & 0.105 \\
Margin             & $-0.197$ & [$-0.296$, $-0.104$] & $-0.011$ & 0.096 \\
Unit price         & $+0.169$ & [$+0.135$, $+0.183$] & $+0.093$ & 0.089 \\
Procure-to-order   & $+0.128$ & [$+0.074$, $+0.173$] & $+0.135$ & 0.105 \\
Customer ID        & $+0.123$ & [$+0.076$, $+0.146$] & $+0.101$ & 0.075 \\
Product ID         & $+0.077$ & [$+0.075$, $+0.079$] & $+0.073$ & 0.069 \\
Discount           & $+0.038$ & [$+0.019$, $+0.058$] & $+0.021$ & 0.029 \\
Order day of week  & $+0.023$ & [$-0.019$, $+0.063$] & $+0.020$ & 0.041 \\
Ship lead time     & $-0.020$ & [$-0.050$, $+0.007$] & $+0.012$ & 0.059 \\
Sales team ID      & $+0.020$ & [$-0.007$, $+0.047$] & $+0.036$ & 0.055 \\
Delivery lead time & $-0.008$ & [$-0.018$, $+0.003$] & $+0.027$ & 0.056 \\
Order month        & $+0.007$ & [$-0.021$, $+0.021$] & $+0.014$ & 0.071 \\
Warehouse          & $+0.006$ & [$-0.035$, $+0.033$] & $+0.004$ & 0.022 \\
Sales channel      & $+0.001$ & [$-0.007$, $+0.006$] & $-0.003$ & 0.017 \\
\bottomrule
\end{tabular}
\end{table}

\begin{figure}[H]
\centering
\includegraphics[width=0.85\linewidth]{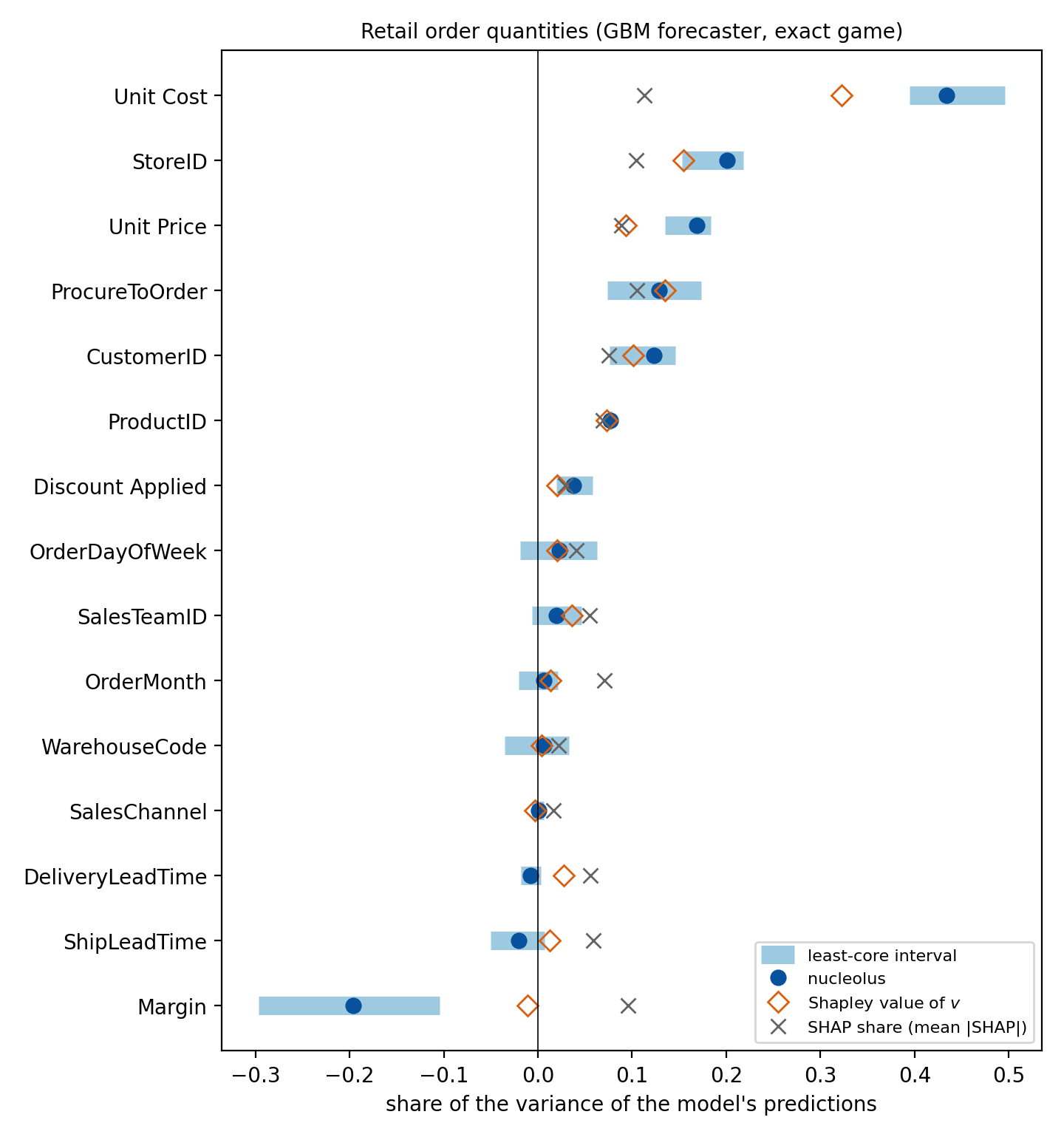}
\caption{Retail forecaster: least-core bands, nucleolus, Shapley value of the
explained-variance game, and mean-$|$TreeSHAP$|$ shares.}
\label{fig:sales}
\end{figure}

Table~\ref{tab:sales} and Figure~\ref{fig:sales} show the results. Here the
nucleolus and the Shapley value of the same game differ substantially (maximum
difference $0.185$; Spearman correlation $0.90$), as they may outside the
2-additive class. The difference is concentrated in the price block. The
nucleolus credits unit cost and unit price with more variance than the Shapley
value does, and gives margin a clearly negative share ($-0.197$ against
$-0.011$). In other words, the stable allocation treats margin as a substitute
whose variation offsets that of cost and price. The Shapley value lies
\emph{outside} the least-core bands for six of the fifteen features. Its worst
coalition is short by $0.790$ of $v(\N)$, and $872$ coalitions have excess above
$\varepsilon^*$. Rescaled TreeSHAP shares are flatter still (maximum excess
$1.036$; $3{,}072$ coalitions above $\varepsilon^*$). The worst-treated coalition
is the same for all three allocations: \{store, product, discount, unit cost,
channel\}. Because the core is empty, no allocation is in the core. The
nucleolus is the allocation that minimises the worst complaint, and by
Proposition~\ref{prop:known}(c) it is certified to belong to the prebargaining
set.

\section{Validation}\label{sec:validation}

\subsection{Faithfulness and complexity}

We measure deletion faithfulness with an AOPC-type score
\citep{samek2017evaluating}: features are set to the baseline in decreasing
order of absolute attribution, and we average the absolute change of the
model's prediction over all deletion steps and 150 instances. Complexity is
measured, following \citet{calzarossa2025assessment}, by the Gini coefficient
of absolute attributions and by the number of features needed for $80\%$ of
the attributed mass.

\begin{table}[H]
\centering
\caption{Faithfulness (AOPC; higher is better) and complexity (Gini; number of
features for $80\%$ of the mass). Random order: mean of 20 random deletion orders.}
\label{tab:faith}
\begin{tabular}{llrrr}
\toprule
Data & Method & AOPC & Gini & $n_{80\%}$ \\
\midrule
Mobile (log price) & Nucleolus ($=$ Shapley of $v$) & 0.980 & 0.616 & 3 \\
                   & SHAP (mean $|\cdot|$)           & 0.980 & 0.556 & 3 \\
                   & LIME (mean $|\cdot|$)           & 0.980 & 0.531 & 3 \\
                   & Random order                     & 0.666 & --- & --- \\
\midrule
Retail (orders)    & Nucleolus                        & 0.412 & 0.587 & 6 \\
                   & Shapley value of $v$             & 0.380 & 0.581 & 6 \\
                   & TreeSHAP (mean $|\cdot|$)        & 0.390 & 0.257 & 10 \\
                   & Random order                     & 0.325 & --- & --- \\
\bottomrule
\end{tabular}
\end{table}

In the mobile model all three methods rank the features identically, so their
deletion curves coincide. They differ in concentration: the variance shares are
the most concentrated. In the retail forecaster the nucleolus ordering is the
most faithful of the three, and it is markedly more concentrated than
TreeSHAP. Deletion faithfulness, however, only reflects the ranking. The
differences highlighted in Section~\ref{sec:illustrations} (bands, objections,
negative shares) concern magnitudes and are invisible to this test.

\subsection{Robustness under resampling}

Following \citet{calzarossa2025assessment}, we re-estimate the mobile model and
recompute its attributions on 50 bootstrap samples. Measured by normalised
absolute shares, two of the seven features have a 5th percentile above the
uninformative $1/7$ for both the nucleolus and mean-$|$SHAP$|$. The mean
5--95\% spread is $0.050$ for the nucleolus and $0.054$ for SHAP, and the mean
Kendall correlation between the bootstrap and full-sample nucleolus rankings
is $0.90$. On average, $74\%$ of bootstrap nucleolus shares fall inside the
full-sample bands. The core is exactly non-empty in $42\%$ of replicates. In the
others, the penalty occasionally admits RAM or screen size with tiny
coefficients of the "wrong" sign, which produces small negative interactions:
the median $\varepsilon^*$ across replicates is $0.0008$ and the maximum
$0.014$. Convexity, and therefore the closed-form bands, is thus robust in
practice, while its exact validity should be checked in each application.

\subsection{Surrogate fidelity at the level of solutions}

\begin{figure}[H]
\centering
\includegraphics[width=\linewidth]{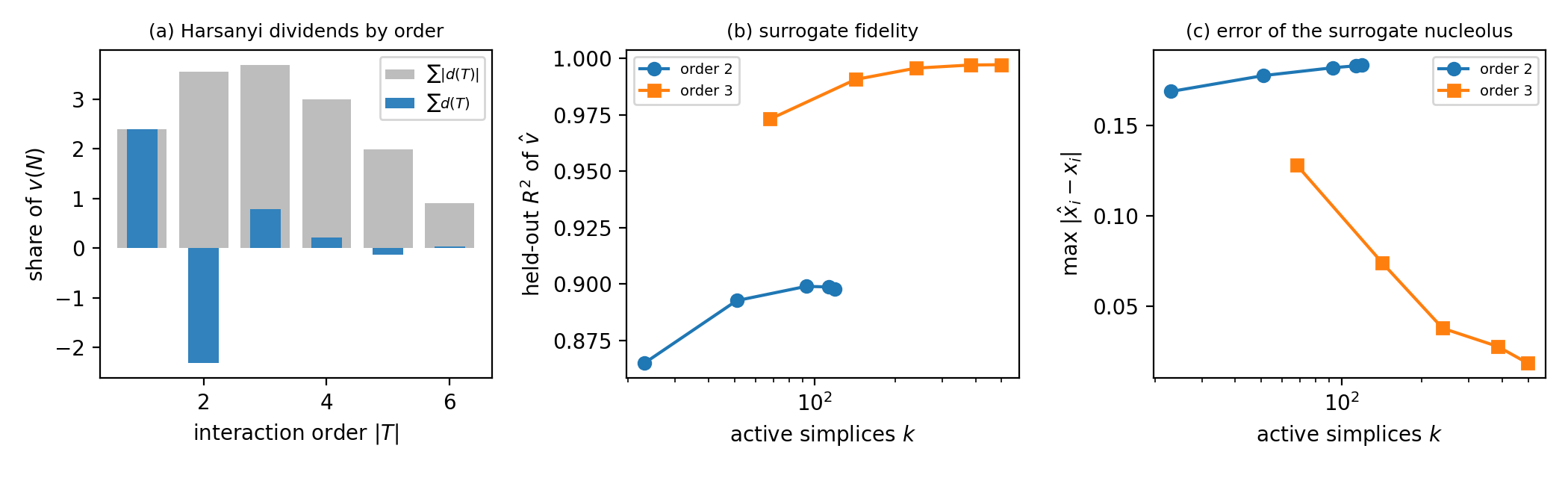}
\caption{Retail forecaster: (a) Harsanyi dividends of the exact game by
interaction order (signed and absolute mass, as shares of $v(\N)$); (b)
held-out fidelity of sparse surrogates; (c) error of the surrogate nucleolus.}
\label{fig:fidelity}
\end{figure}

The exact retail game lets us evaluate the sparse surrogates of
Section~\ref{sec:surrogates} against the truth. We fit 2- and 3-additive LASSO
surrogates on $2{,}000$ random coalitions and evaluate them on the remaining
$30{,}766$ (Figure~\ref{fig:fidelity}).

\begin{table}[H]
\centering
\caption{Retail forecaster: sparse $k$-additive surrogates fitted on 2,000
coalitions. $R^2$ on the 30,766 held-out coalitions; error of the surrogate
nucleolus relative to the exact one.}
\label{tab:fidelity}
\begin{tabular}{rrrrrc}
\toprule
Order & Active dividends & Held-out $R^2$ & $\max_i|\hat x_i-x_i|$ & Spearman & Top-5 overlap \\
\midrule
2 & 23  & 0.865 & 0.169 & 0.914 & 3/5 \\
2 & 51  & 0.893 & 0.178 & 0.914 & 3/5 \\
2 & 119 & 0.898 & 0.184 & 0.914 & 4/5 \\
3 & 68  & 0.973 & 0.128 & 0.882 & 4/5 \\
3 & 142 & 0.991 & 0.074 & 0.950 & 5/5 \\
3 & 239 & 0.996 & 0.038 & 0.971 & 5/5 \\
3 & 498 & 0.997 & 0.018 & 0.989 & 5/5 \\
\bottomrule
\end{tabular}
\end{table}

The pairwise surrogates illustrate Theorem~\ref{thm:2add} at work. Their fit
to $v$ is respectable ($R^2\approx0.90$), but their nucleolus is their Shapley
value, so it cannot track the exact nucleolus, and adding pairwise terms does
not help (the error stays at $0.17$--$0.18$). Third-order surrogates recover
the exact nucleolus to within $0.04$ with 239 active dividends and $0.02$ with
498. Game-level fidelity is therefore a poor guide to solution-level fidelity,
and whenever stability concepts are the object of interest the surrogate's
interaction order must be chosen with that in mind.

\subsection{Forcing convexity on the surrogate}

A tempting shortcut is to fit the surrogate with non-negative interaction
dividends. The surrogate game is then convex, its core is non-empty, and every
reported allocation can be certified with one minimum cut
(Proposition~\ref{prop:mincut}). We tested this on the exact retail game, using
the same 2,000 training coalitions, interaction orders 2 and 3, and penalties
$\alpha\in\{10^{-3},10^{-4}\}$.

The result is the same in all four cases. The constraint removes every
interaction term, so the fitted surrogate is additive: only the 15 singleton
dividends remain, and this persists without any penalty ($\alpha=0$). Its
held-out $R^2$ falls to $0.32$, against $0.86$--$0.997$ for the unconstrained
surrogates of Table~\ref{tab:fidelity}. An additive game has a one-point core
equal to its nucleolus, so the minimum-cut certificate is satisfied (maximum
excess $<10^{-14}$) and the allocation is reported as stable. In the true
game, however, that allocation leaves a coalition short by $0.715$--$0.716$ of
$v(\N)$, above the least-core value $\varepsilon^*=0.608$, and 36 to 43
coalitions exceed $\varepsilon^*$. Its distance to the exact nucleolus is
$0.17$. The certificate is therefore exact for the surrogate and uninformative
about the model. Where non-negativity holds in the true game, as in the linear
mobile model, the closed-form bands and certificates are valid without
constraining anything; where it does not, the honest report is the least-core
band of the actual game.

\subsection{A group-comparison diagnostic}

\citet{chen2024measuring} assess fairness by comparing attributions across
groups. To illustrate the mechanism, we form for each test-period order the
score $s(x)=\sum_i\bar x_i z_i(x)$, where $\bar x$ is the nucleolus normalised
by $\sum_i|x_i|$ and $z_i$ the standardised feature, and we compare its
distribution across the four sales channels with pairwise Kolmogorov--Smirnov
tests. None of the six comparisons is significant at $5\%$ (largest statistic
$0.069$, Distributor vs.\ In-Store, $p=0.105$). Sales channel is not a
protected attribute, and this exercise only shows how stability-based
attributions plug into such diagnostics. A substantive audit requires a
genuine protected attribute and a model with predictive skill.

\section{Discussion}\label{sec:discussion}

\paragraph{What stability adds.} For linear models, and for any explanation
built on a pairwise surrogate, the nucleolus coincides with the Shapley value
of the game being explained (Theorem~\ref{thm:2add}). In that class the
distinctive contribution of stability concepts is not a different point but
(i) the bands, which quantify how much of each feature's credit is
determined by the data and how much is a matter of how jointly produced
variance is split, and (ii) certificates, which reveal that common global
explanations, such as rescaled mean-$|$SHAP$|$ or LIME shares, can be
contested by an identifiable coalition. When the model has higher-order
interactions, as in the retail forecaster, the nucleolus becomes a genuinely
different point, and the bands can exclude the Shapley value.

\paragraph{Choice of game.} All results are conditional on the game. We use
an interventional baseline and the variance of predictions. Other choices,
such as conditional expectations \citep{sundararajan2020many}, loss-based
worths \citep{covert2020understanding} or local games around a single
prediction, yield different games, to which the same machinery applies.
Negative shares and empty cores are features of the explained-variance game
under substitution, not artefacts of the algorithms.

\paragraph{Limitations.} Exact computation requires evaluating $v$ on all
coalitions, which is feasible up to roughly 20 features. Beyond that we rely on
sparse surrogates, whose order must be validated at the solution level, as in
Table~\ref{tab:fidelity}. The exact objection test is exponential and is
practical only for small $n$; for larger games we certify the nucleolus
through Proposition~\ref{prop:known}(c). Finally, explanations of a model with
no predictive skill, like our retail forecaster, explain the model and nothing
more.

\paragraph{Other uses.} The same construction applies when the players are
training observations rather than features (data valuation), and to groups of
features, for example a variable together with all its lags in a time-series
forecaster. Grouping reduces $n$ and keeps explanations economically
interpretable.

\section{Conclusion}\label{sec:conclusion}

We have proposed to report feature attributions as stability bands, with the
nucleolus as a distinguished point, for an explained-variance game whose
worths are shares of the variance of a model's predictions. For linear models
this game is exactly 2-additive and its interactions are generated by
multicollinearity. For every 2-additive game the prenucleolus is the Shapley
value, the core is symmetric around it, and core emptiness is a cut condition
on the interaction graph. For convex games the bands are explicit and any
proposed explanation can be certified with a single minimum cut. The two
applications show both sides of these results. In a convex linear model the
Shapley decomposition is the stable centre, while common global SHAP and LIME
explanations admit justified objections. In a non-additive forecaster, the
stable explanation differs from the Shapley value, and pairwise surrogates
cannot recover it. Natural extensions are local and loss-based games, grouped
players for time-series models, adaptive choice of the surrogate's order, and
fairness diagnostics with genuine protected attributes.

\bigskip
\noindent\textbf{Code and data availability.} Both datasets are publicly
available on Kaggle (\url{https://bit.ly/4kCyJ3O};
\url{https://www.kaggle.com/datasets/talhabu/us-regional-sales-data}). The code
reproducing every number, table and figure (\nolinkurl{code/cgt_analysis.py},
\nolinkurl{code/cgt_solvers.py}) accompanies the paper, together with regression
tests (\nolinkurl{code/tests/test_reviewer_examples.py}).

\noindent\textbf{Conflict of interest.} The authors declare no conflict of interest.

\noindent\textbf{Funding.} This project was supported by PGI-UNS 24E172.

\noindent\textbf{CRediT author statement.} Fernando Delbianco:
Conceptualization, Investigation, Software, Funding acquisition, Methodology,
Writing -- original draft. Fernando Tohm\'e: Investigation, Formal analysis,
Funding acquisition, Writing -- original draft.

\bibliographystyle{apalike}
\bibliography{references}

\end{document}